\documentclass[journal]{IEEEtran}
\usepackage{amsmath,amssymb,amscd,latexsym,dsfont}
\usepackage{comment}
\usepackage{color}
\usepackage{subfigure}
\usepackage{enumerate}
\usepackage{enumitem}
\usepackage{stfloats}
\usepackage{psfrag}
\usepackage{setspace}
\usepackage{cite}
\usepackage{balance}
\usepackage{enumitem}
\usepackage{tikz}
\usepackage{pgfplots,amsfonts}
\usepackage{pgfplotstable}
\usetikzlibrary{shapes}
\usetikzlibrary{spy}
\usetikzlibrary{fit}
\usetikzlibrary{shapes.multipart}
\usetikzlibrary{positioning}

\usetikzlibrary{plotmarks,matrix,chains,scopes,fit,calc,shapes,positioning,decorations,intersections,arrows,backgrounds,shadows}

\newlength\FigureHeight
\newlength\FigureWidth
\definecolor{myDarkGreen}{rgb}{0.00000,0.58824,0.00000}%
\definecolor{uniform}{rgb}{0.00000,0.58824,0.00000}%
\definecolor{AWGNreference}{rgb}{0.00000,0.58824,0.00000}%

\newcounter{lemma}

\newtheorem{exampleplain}{Example}

\newtheorem{definitionplain}{Definition}

\newcommand{\bc}{\boldsymbol{c}}
\newcommand{\bb}{\boldsymbol{b}}
\newcommand{\bX}{\boldsymbol{X}}\newcommand{\bx}{\boldsymbol{x}}

\newcommand{\bs}{\boldsymbol{s}}

\usepackage[utf8]{inputenc}

\usepackage{amsmath,amssymb}

\usepackage{xcolor}
\usepackage{graphicx}
\usepackage{tikz}
\usepackage{pgfplots}
\usepackage{float}
\usepackage{balance}

\usepackage{xspace}
\usepackage[normalem]{ulem}

\usepackage{dsfont,empheq}
\usepackage{multicol,multirow}

\usepackage{cite}
\usepackage{gensymb}

\usepackage{booktabs, cellspace, hhline}

\pgfplotsset{compat=1.14}
\usepackage{wrapfig}
\usepackage[acronym,nomain]{glossaries}

\newacronym{CUT}{CUT}{channel under test}

\newacronym{KK}{KK}{Kramers-Kronig}
\newacronym{CSPR}{CSPR}{carrier-to-signal power ratio}
\newacronym{KKRX}{KKRx}{Kramers-Kronig Receiver}
\newacronym{SSBI}{SSBI}{signal-signal beat interference}

\newacronym{GS}{GS}{geometric shaping}
\newacronym{PS}{PS}{probabilistic shaping}
\newacronym{DSP}{DSP}{digital signal processing}
\newacronym{MIMO}{MIMO}{multiple-input multiple-output}
\newacronym{TDE}{TDE}{time domain equalizer}
\newacronym{FDE}{FDE}{frequency domain equalizer}
\newacronym{LMS}{LMS}{least means square}
\newacronym{DDLMS}{DD-LMS}{decision directed least means square}
\newacronym{FFE}{FFE}{feed-forward equalizer}
\newacronym{FBE}{FBE}{feedback equalizer}
\newacronym{BPS}{BPS}{blind phase search}

\newacronym{SMF}{SMF}{single-mode fiber}
\newacronym[plural=SSMFs]{SSMF}{SSMF}{standard single-mode fiber}
\newacronym[plural=FMFs]{FMF}{FMF}{few-mode fiber}
\newacronym{FMF12}{FMF12}{12 mode FMF}
\newacronym{MMF}{MMF}{multi-mode fiber}
\newacronym{SI}{SI}{step index}
\newacronym{GI}{GI}{graded index}
\newacronym{DCF}{DCF}{dispersion compensated fiber}

\newacronym{SDM}{SDM}{space division multiplexing}
\newacronym{MDM}{MDM}{mode division multiplexed}
\newacronym{WDM}{WDM}{wavelength division multiplexing}
\newacronym{DWDM}{DWDM}{dense wavelength division multiplexing}
\newacronym{LP}{LP}{linear polarized}
\newacronym[plural=MMUXs,firstplural=mode multiplexers]{MMUX}{MMUX}{mode multiplexer}
\newacronym{PL}{PL}{photonic lantern}
\newacronym{3DWG}{3DWG}{3D-waveguide}
\newacronym{MDL}{MDL}{mode dependent loss}
\newacronym{DGD}{DGD}{differential group delay}
\newacronym{DMGD}{DMGD}{differential mode group delay}
\newacronym{QSM}{QSM}{quasi-single-mode}
\newacronym{GIMMF}{GI-MMF}{graded-index multi-mode fiber}

\newacronym{SSB}{SSB}{single side band}
\newacronym{QPSK}{QPSK}{quadrature phase shift keying}
\newacronym{QAM}{QAM}{quadrature amplitude modulation}
\newacronym{RRC}{RRC}{root-raised-cosine}
\newacronym{4D-64PRS}{4D-64PRS}{
four-dimensional 64-ary polarization-ring-switching}
\newacronym{DP}{DP}{dual-polarization}
\newacronym[\glslongpluralkey=states-of-polarization]{SOP}{SOP}{state-of-polarization}
\newacronym{PM}{PM}{polarization-multiplexed}

\newacronym{ECL}{ECL}{external cavity laser}
\newacronym{CW}{CW}{continuous wave}
\newacronym[plural=DFBs]{DFB}{DFB}{distributed feedback laser}

\newacronym[plural=DACs]{DAC}{DAC}{digital-to-analog converter}
\newacronym{ADC}{ADC}{analog-to-digital converter}
\newacronym{PRBS}{PRBS}{pseudo-random bit sequence}
\newacronym{LO}{LO}{local oscillator}
\newacronym{EDFA}{EDFA}{erbium-doped fiber amplifier}
\newacronym{MZM}{MZM}{Mach-Zehnder modulator}
\newacronym{DP-MZM}{DP-MZM}{dual-polarization Mach-Zehnder modulator}
\newacronym{ChUT}{ChUT}{channel under test}
\newacronym{WSS}{WSS}{wavelength selective switch}
\newacronym[plural=VOAs]{VOA}{VOA}{variable optical attenuator}
\newacronym[plural=PDCRXs]{PDCRX}{PDCRX}{polarization diverse coherent receiver}
\newacronym{DSO}{DSO}{digital storage oscilloscope}
\newacronym{ASE}{ASE}{amplified spontaneous emission}
\newacronym{PBS}{PBS}{polarization beam splitter}
\newacronym{PD}{PD}{photodiode}
\newacronym{AOM}{AOM}{acousto-optic modulator}
\newacronym{BPD}{BPD}{balanced photo-diode}
\newacronym{OMFT}{OMFT}{optical multi-format transmitter}
\newacronym{DPIQ}{DP-IQM}{dual-polarization IQ-modulator}
\newacronym{ABC}{ABC}{automatic bias controller}
\newacronym{OTF}{OTF}{optical tunable filter}
\newacronym{LSPS}{LSPS}{loop-synchronous polarization scrambler}
\newacronym{OSA}{OSA}{optical spectrum analyzer}

\newacronym{OSNR}{OSNR}{optical signal to noise ratio}
\newacronym{BER}{BER}{bit error rate}
\newacronym{IL}{IL}{insertion loss}

\newacronym{SDFEC}{SD-FEC}{soft-decision forward error correction}
\newacronym{HDFEC}{HD-FEC}{hard-decision forward error correction}
\newacronym{FEC}{FEC}{forward error correction}
\newacronym{LDPC}{LDPC}{low-density parity-check code}

\newacronym{AIR}{AIR}{achievable information rate}
\newacronym{AR}{AR}{achievable rates}
\newacronym{MI}{MI}{mutual information}
\newacronym{GMI}{GMI}{generalized mutual information}
\newacronym{BICM}{BICM}{bit-interleaved coded modulation}

\newacronym{OVNA}{OVNA}{optical vector network analyzer}
\newacronym{NIR}{NIR}{near infrared}
\newacronym{CD}{CD}{chromatic dispersion}
\newacronym{OTDR}{OTDR}{optical time domain reflectometry}
\newacronym{OFDR}{OFDR}{optical frequency domain reflectometry}
\newacronym{GPU}{GPU}{graphics processing unit}
\newacronym{SVD}{SVD}{singular value decomposition}
\newacronym{WGN}{WGN}{white Gaussian noise}
\newacronym{AWGN}{AWGN}{additive white Gaussian noise}
\newacronym{PDL}{PDL}{polarization dependent loss}
\newacronym{SPS}{sps}{samples-per-symbol}
\newacronym{SE}{SE}{spectral efficiency}

\usepackage{arydshln}

\usetikzlibrary{arrows, shapes.gates.logic.US, calc}
\tikzstyle{branch}=[fill, shape=circle, minimum size=3pt, inner sep=0pt]

\IEEEoverridecommandlockouts

\title{Low-Complexity Geometrical Shaping for 4D Modulation Formats via Amplitude Coding}
\author{Bin~Chen,~\IEEEmembership{Member,~IEEE}, Wei Ling, Yunus Can G\"ultekin,~\IEEEmembership{Member,~IEEE}, Yi Lei, \\Chigo Okonkwo,~\IEEEmembership{Senior Member,~IEEE}, 
and Alex~Alvarado,~\IEEEmembership{Senior Member,~IEEE}
\thanks{B. Chen, W. Ling, and Y. Lei are with the School of Computer Science and Information Engineering, Hefei University of Technology, Hefei, China. (e-mails:~\{bin.chen,leiyi\}@hfut.edu.cn, 18226958877@163.com).}
\thanks{Y. C. G\"ultekin, C. Okonkwo and A. Alvarado are with the Department of Electrical Engineering, Eindhoven University of Technology, 5600 MB, Eindhoven, The Netherlands (e-mails:~\{y.c.g.gultekin,a.alvarado,cokonkwo\}@tue.nl).}

\thanks{
The work is partially supported by the National Natural Science Foundation of China (NSFC) under Grant  (62171175,  62001151), Fundamental Research Funds for the Central Universities under Grant JZ2020HGTB0015 and the Netherlands Organisation for Scientific Research (NWO) via the VIDI Grant ICONIC (project number 15685). }
}

\begin{document}
\bstctlcite{IEEEexample:BSTcontrol}

\maketitle

\begin{abstract}
Signal shaping is vital to approach Shannon’s capacity, yet it is challenging to implement at very high speeds. For example, probabilistic shaping often requires arithmetic coding to realize the target distribution. Geometric shaping 
requires look-up tables to store the constellation points. 
In this paper, we propose a four-dimensional amplitude coding (4D-AC) geometrical shaper architecture. The proposed architecture can generate in real time geometrically shaped 4D formats via simple logic circuit operations and two conventional quadrature amplitude modulation (QAM) modulators. This paper describes the 4D-AC  used in generating approximated versions of two recently proposed 4D orthant symmetric modulation formats with spectral efficiencies of 6~bit/4D-sym and 7~bit/4D-sym, {respectively}. Numerical results show losses below $0.05$~dB when compared against the baseline formats.
\end{abstract}

\section{Introduction}

The combination of high-order modulation formats and forward error correction (FEC)---known as coded modulation (CM)---is indispensable for digital communication systems targeting high transmission rates. Among the different CM schemes, such as multilevel coding
(MLC) \cite{
WachsmannTIT1999}, trellis CM \cite{UngerboeckTIT1982}, and bit-interleaved coded modulation (BICM) \cite{
Szczecinski2015BICM}, BICM has attracted a lot of attention due to its low complexity.

Constellation shaping has been widely investigated in optical fibre communications via probabilistic shaping (PS) and geometric shaping (GS). PS and GS has been shown to provide linear shaping gains and also to provide tolerance against nonlinearities. The performance of PS has been examined in theory, simulations, and experiments \cite{BochererTCOM2015,TobiasJLT16,Buchali2016,RennerJLT2017}. 
GS refers to a nonrectangular and often a multidimensional (MD) modulation format, which
 has been studied in optical communications in the context of 1D \cite{SillekensECOC2018} and 2D \cite{RasmusECOC2018,BinECOC2018,IonescuJLT2020} GS optimizations. 
GS is also currently implemented in commercial products \cite{FujitsuT600}.
Hybrid shaping approaches in which GS and PS are combined have been investigated to approach a Gaussian distribution with limited number of constellation points \cite{8346117}.


In order to obtain performance gains by using an MD signal space, modulation formats have been investigated in optical communications for both linear noise tolerannce \cite{Karlsson:09, MiraniJLT2020,LiArxiv2021} and nonlinear resilience \cite{RasmusECOC2018,El-RahmanJLT2018,BinChenJLT2019}. MD formats with constant modulus \cite{ReimerOFC2016,Kojima2017JLT,BinChenJLT2019} and/or polarization balancing properties \cite{ReimerOFC2016,Bendimerad:18,BinChenPTL2019} are able to mitigate the optical fiber nonlinear interference. Four-dimensional (4D) formats are also currently commercially available\cite{FujitsuT600}. 4D orthant symmetric (OS) modulation formats have been recently proposed \cite{BinChenJLT2019,ChenJLT2021}. The formats in \cite{BinChenJLT2019,ChenJLT2021} have spectral efficiencies of 6~bit/4D-sym and 7~bit/4D-sym, and are called 4D-64PRS and 4D-OS128, {respectively}. 4D-64PRS is a constant-modulus format. 4D-OS128 is not constant modulus but it has low symbol energy variations, and thus, it is resilient to nonlinearities.

The operation principle behind MD modulation formats is straightforward, however, a key aspect in its implementation is the complexity of their mapper and demapper. Designing low-complexity mappers and demappers suitable for real-time processing is challenging. For high-order MD modulation formats, the MD mapper requires large look-up tables (LUTs) for storing the constellation points, while the MD demapper for soft-decision (SD) decoders requires the calculation of Euclidean distances (ED) between the MD received signal and all symbols of the MD constellation. 
An alternative to storing large LUTs is to generate the constellation points by subset-optimization and set-partitioning from conventional modulations \cite{SjodinCOML2013,Bendimerad:18,InwoongOFC2019} or lattice-based modulations \cite{MiraniJLT2020}.

In this work, a novel 4D mapping architecture is proposed for on-the-fly generation of 4D modulation formats. We call the proposed architecture four-dimensional amplitude coding (4D-AC). The proposed solution selects a subset of polarization-multiplexed 16-ary quadrature amplitude modulation (PM-16QAM) symbols via logical operations. This logic circuit is to be included between the FEC and the PM-16QAM mapper, and thus, our proposed architecture can be easily implemented as an \textit{add-on} to existing PM-16QAM systems. 
We show that the proposed 4D-AC  can be used to generate approximated versions of 4D-64PRS \cite{BinChenJLT2019} and 4D-OS128\cite{ChenJLT2021}, {respectively}. Approximation losses below $0.05$~dB are reported. {Thus, the proposed 4D-AC  can keep the advantages of 4D-64PRS  and  4D-OS128  formats, with no need for LUTs.}

\section{Multidimensional Amplitude Coding}
\subsection{4D-AC  with Four 4ASK or Two 16QAM Mappers}
\begin{figure*}[!tb]
  \centering
 \scalebox{0.93}{\includegraphics[]{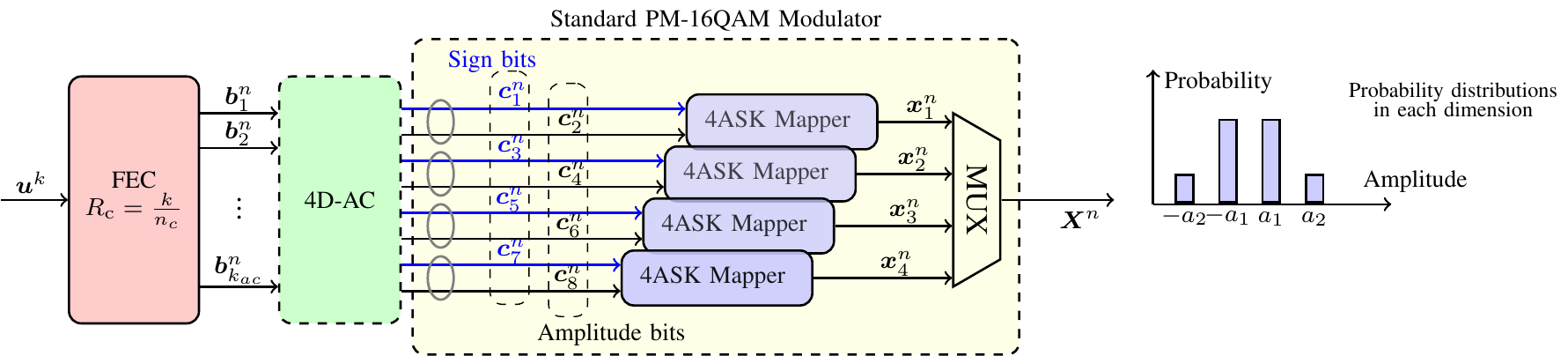}}
 \vspace{-0.5em}
  \caption{Signal shaping via four-dimensional amplitude coding (4D-AC).  An example of probability distribution for each 1D is shown at the output of modulator.
}
  \label{fig:structure}
\vspace{-1em}
\end{figure*}

Fig. \ref{fig:structure} shows the transmitter structure under consideration, which represents a generic CM system consisting of a FEC encoder followed by a MD  mapper.
We consider the most popular case of 4-dimensional (4D) transmission in fiber optical links: coherent communications using two polarizations of the light  (4 real dimensions: $I_x, Q_x, I_y, Q_y$).
A standard PM-$2^{2m}$QAM system transmits 4D symbol sequences $\bX^n=[\bx^n_1; \bx^n_2; \bx^n_3;\bx^n_4]$  of length $n$, where each 4D symbol $\bx=(x_1,x_2,x_{3},x_{4})\in\mathcal{X}^4$ is drawn from the 4-fold Cartesian product of the $2^m$ASK alphabet $\mathcal{X}=\{\pm a_1,\pm a_2,\dots,\pm a_{2^{m-1}}\}$.
Furthermore, this 1D alphabet can be factorized as $\mathcal{X}=\mathcal{S}\mathcal{A}$, where $\mathcal{S}=\{+1, -1\}$ and $\mathcal{A}=\{a_1, a_2,\dots,a_{2^{m-1}}\}$ are the sign and amplitude alphabets, respectively. 
Each 4D symbol $\bx=(x_1,x_2,x_{3},x_{4})$ is labeled by a length-$4m$ binary sequence $\bc=(c_1,c_2\dots,c_{4m})\in\{0,1\}^{4m}$.
In this paper, the binary reflected Gray code (BRGC) is used for labeling. For the case of $m=2$, the amplitude alphabet $[-a_2, -a_1, a_1, a_2]$ (4ASK) is then labeled by $[11, 10, 00, 01]$.

In Fig. \ref{fig:structure}, a $k$-bit information sequence is first encoded by a rate $R_c=k/n_c$ FEC code, where $n_c=nk_{ac}$.
Then, the coded sequence $\bb^{n_c}$ is divided into $k_{ac}$ parallel streams $\bb^n_1,\bb^n_2,\dots,\bb^n_{k_{ac}}$.
 Next, a 4D-AC block (green) \textit{encodes} sequences of $k_{ac}$ bits into an $8$-bit string, creating 8 parallel streams $\bc^n_1,\bc^n_2,\dots,\bc^n_{8}$.
 Finally, four 4ASK mappers (or equivalently two 16QAM mappers) transform these $8$ bits into four $4$ASK symbols (or two 16QAM symbols).
 The $k$-bit input sequence is therefore mapped to an $n$-tuple of 4D symbols and the resulting net information rate is $k/n$~bit/4D-sym.
 The encoding procedure in the 4D-AC block is designed such that its output, as well as the resulting 4D symbols in each 1D projection (see an example of the modulator output in right side of Fig. \ref{fig:structure}) have a nonuniform distribution.
 
The rate of the 4D-AC in Fig. \ref{fig:structure} is $k_{ac}/8$. 
In the absence of the 4D-AC block (green), the system is equivalent to a standard PM-16QAM transmitter.
In general, the 4D-AC can be designed to generate 4D modulation format with SE of up to $4m$~bit/4D-sym for an existing PM-$2^{2m}$QAM system.
Our objective is to design the 4D-AC block targeting spectral efficiencies below $4m$~bit/4D-sym and at the same time obtaining shaping gain.
In Sec.~II-B and Sec.~II-C, we will provide the design of 4D-AC to generate 4D formats with SEs of 6~bit/4D-sym and 7~bit/4D-sym.
In these cases, $2^6=64$ and $2^7=128$ 4D points out of the possible $(2^2)^4=256$ PM-16QAM symbols will be selected.

\subsection{4D-AC  for 6 bit/4D-sym (6b4D-AC)}

\begin{figure}[!tb]
\centering
 \scalebox{0.52}{\includegraphics[]{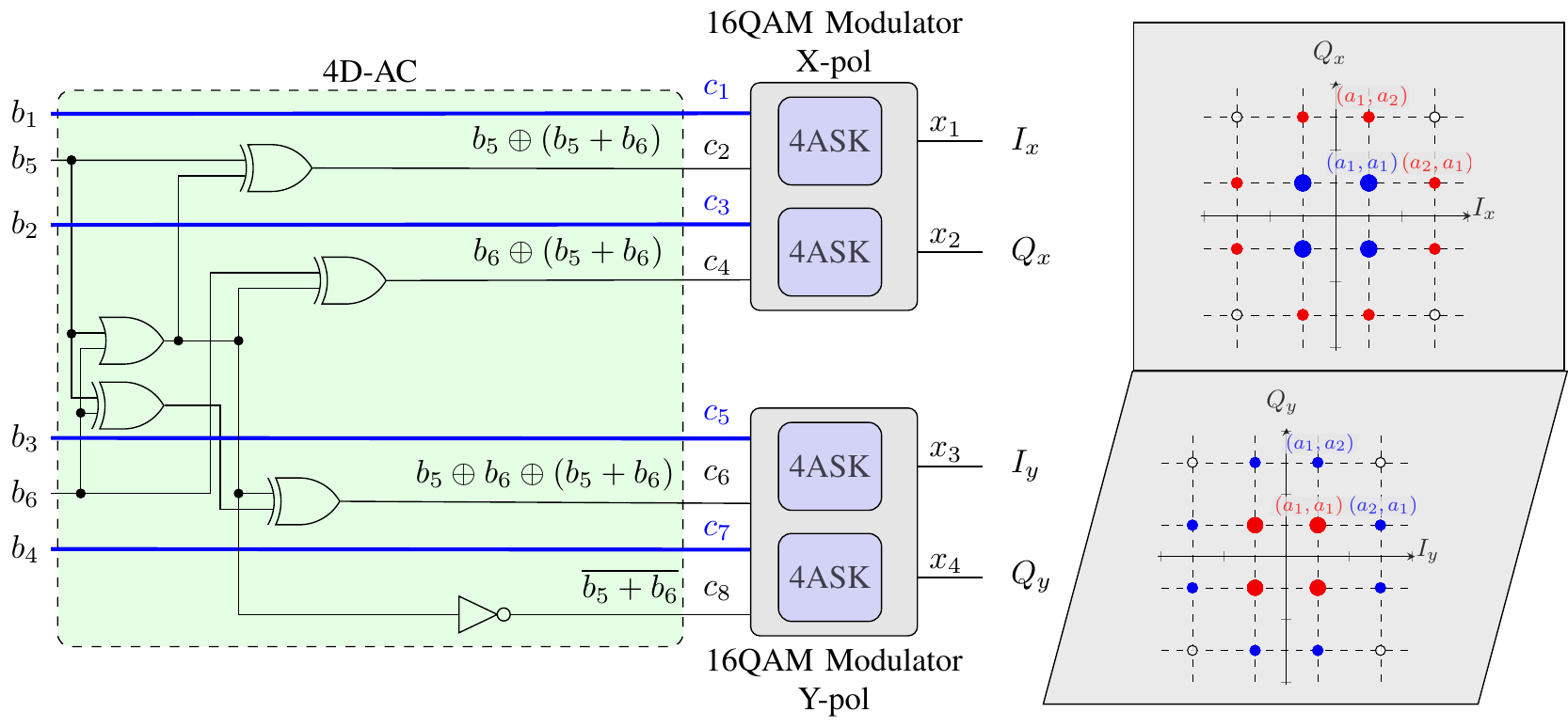}}
\vspace{-1.5em}
\caption{6~bit/4D-sym 4D-AC  for generating non-uniform probability distribution with using PM-16QAM transmitter. 
In 4D projection, there are 64 4D-points with equal probability $P=1/64$. In each 2D projection, white points: $P=0$, large colored points: $P=8/64$, small colored points: $P=4/64$.}
\label{fig:mapper_6bit_v2}
\vspace{-1em}
\end{figure}

Here we explain how to realize transmission at SE of 6~bit/4D-sym by generating the recently proposed 4D modulation format 4D-64PRS \cite{BinChenJLT2019}. As shown in Fig.~\ref{fig:mapper_6bit_v2}, we achieve this by using $k_{ac}=6$.
Through four conventional 4ASK mappers, these 256 points are labeled with 8-bit vectors, four of which are sign bits, and where the remaining are amplitude bits.
Out of the $k_{ac}=6$ bits at the input of the 4D-AC, four are assigned as sign bits without change (highlighted as blue in Fig.~\ref{fig:mapper_6bit_v2}), i.e.,
\begin{align}\label{eq:nonlinearcoding0}
c_1=b_1; \ \ c_3=b_2;\ \ c_5=b_3;  \ \  c_7=b_4.
\end{align}

Due to the geometry and the labeling of the 4D-64PRS format, the bits that select the amplitudes (for each real dimension) through four 4ASK mappers are not uniform.
 These four amplitude bits ($c_2, c_4, c_6$ and $c_8$, see Fig.~\ref{fig:mapper_6bit_v2}) can be obtained through the following formulas:
\begin{align}\label{eq:nonlinearcoding}\nonumber
&c_2={b_5\oplus {(b_5+b_6)}}; \ \ \ \ \ \ \ \ \ \ c_4={b_6\oplus {(b_5+b_6)}};\\ 
&c_6={b_5\oplus b_6\oplus{(b_5+b_6)}}; \ \ \ \ c_8=\overline{b_5+b_6}.
\end{align}
In \eqref{eq:nonlinearcoding}, ``$\oplus$'' denotes the logical XOR operation, while ``$+$'' and ``$\overline{(\cdot)}$" denote logical OR and NOT, {respectively}.
The formulas in \eqref{eq:nonlinearcoding} are examples of nonlinear coding reflecting nonlinear constraints on the bits. 

Fig. \ref{fig:mapper_6bit_v2} (left) shows a block diagram of a logical circuit that implements \eqref{eq:nonlinearcoding0}--\eqref{eq:nonlinearcoding}.
Every 8-bit 4D-AC outputs are fed to two standard 16QAM modulators, which output a 4D symbol $\bx=(x_1, x_2, x_3, x_4)=\boldsymbol{\alpha}\circ \bs$, where $\circ$ denotes Hadamard product, $\alpha_i\in\mathcal{A}= \{a_1,a_2\}$ and $s_i\in\mathcal{S}=\{-1, +1\}$ denote the amplitude and sign of the 4D symbol in $i$th dimension, respectively. 
Depending on \eqref{eq:nonlinearcoding}, the mapping operations $[b_5b_6]\longmapsto[c_2c_4c_6c_8]\longmapsto \boldsymbol{\alpha}$ and $[b_1b_2b_3b_4]\longmapsto[c_1c_3c_5c_7]\longmapsto \bs$ are as follows, 
  \begin{itemize}
    \item $[00]\longmapsto[0001]\longmapsto(\pm a_1,\pm a_1,\pm a_1,\pm a_2)$
    \item $[01]\longmapsto[1000]\longmapsto(\pm a_2,\pm a_1,\pm a_1,\pm a_1)$
    \item $[11]\longmapsto[0010]\longmapsto(\pm a_1,\pm a_1,\pm a_2,\pm a_1)$
    \item $[10]\longmapsto[0100]\longmapsto(\pm a_1,\pm a_2,\pm a_1,\pm a_1)$.
  \end{itemize}
 The sign ``$\pm $" of these four output sequences are decided by the partial  output (sign bits) of 4D-AC  $[c_1c_3c_5c_7]\longmapsto\bs=\left((-1)^{c_1},(-1)^{c_3},(-1)^{c_5},(-1)^{c_7}\right)\in\{-1,+1\}^4$. 
 
In order to clearly show the inter-polarization dependency of the 4D shaped modulation format in Fig. \ref{fig:mapper_6bit_v2} (right), the two 2D projections are shown via a color
coding strategy: 2D projected symbols in the X-pol and Y-pol are valid 4D symbols only if they share the same color (red or blue).

\subsection{4D-AC for 7~bit/4D-sym (7b4D-AC)}
\begin{figure}[!tb]
\centering
\scalebox{0.52}{\includegraphics[]{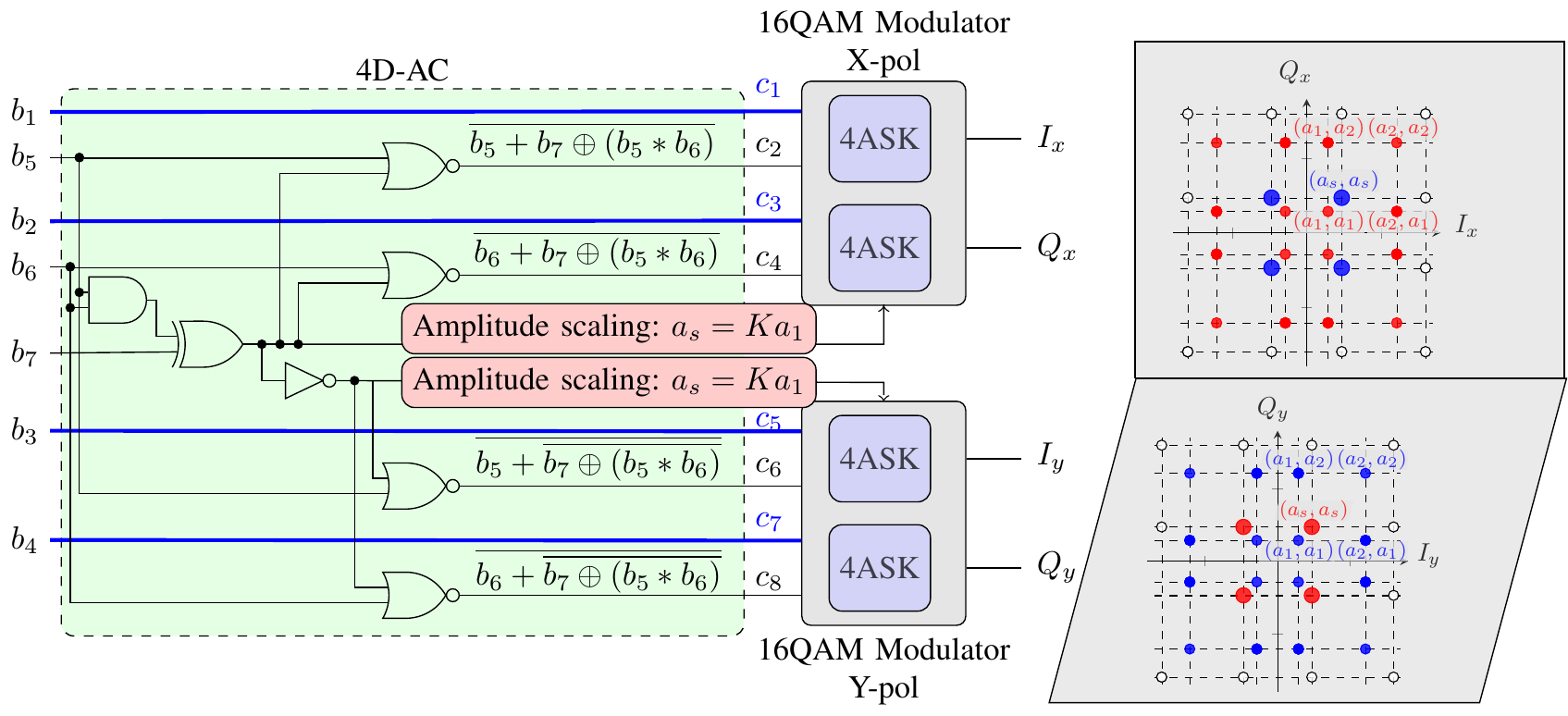}}
\vspace{-1.5em}
\caption{7~bit/4D-sym 4D-AC for generating non-uniform probability distribution with using PM-16QAM transmitter. 
In 4D projection, there are 128 4D-point with equal probability $P=1/128$. In each 2D projection, white points: $P=0$,  large colored points: $P=16/128$, small colored points: $P=4/128$.}
\label{fig:mapper_7bit_v2}
\vspace{-1em}
\end{figure}

Similar to the MD-AC approach for 6~bit/4D-sym modulation, 7~bit/4D-sym modulation {4D-OS128} \cite{ChenJLT2021} can also be generated by using PM-16QAM structure, which selects 128 of 256 PM-16QAM symbols with a design modification (amplitude scaling). The 4D-AC  for generating 7~bit modulation is  shown in Fig. \ref{fig:mapper_7bit_v2}.
Due to the structure of 4D-OS128, two out of four amplitudes of a 4D symbol  should be scaled differently than the others.
Therefore, we consider 
the  amplitude alphabets $\mathcal{A}=\{a_1,a_s,a_2\}$ for $a_1>0, a_s=Ka_1>0, a_2>0$, where $K$ is an amplitude scaling factor.\footnote{{Note that the scaling operation will require a  higher resolution  for analog-to-digital/digital-to-analog  converter with respect to PM-16QAM.}} The sign alphabet is $\mathcal{S}=\{-1, +1\}$ and the channel input is denoted as $\bx=(x_1, x_2, x_3, x_4)=\boldsymbol{\alpha} \circ \bs$.
The four sign bits $c_1, c_3, c_5, c_7$ (highlighted as blue in Fig.~\ref{fig:mapper_7bit_v2}) are directly assigned from the first four bits of the $k_{ac}= 7$~bits at the input of the 4D-AC.
Four amplitude bits $c_2, c_4, c_6, c_8$ are now obtained through the following formulas:
\begin{align}\label{eq:nonlinearcoding7bit}\nonumber
&c_2=\overline{b_5+{b_7\oplus {(b_5*b_6)}}}; \ \ &c_4=\overline{b_6+{b_7\oplus {(b_5*b_6)}}};\\ 
&c_6=\overline{b_5+\overline{b_7\oplus {(b_5*b_6)}}}; \ \ &c_8=\overline{b_6+\overline{b_7\oplus {(b_5*b_6)}}},
\end{align}
where $``*"$ denotes the logical AND operation.

  Depending on \eqref{eq:nonlinearcoding7bit}, the mapping operations $[b_5b_6b_7]\longmapsto[c_2c_4c_6c_8]\longmapsto \boldsymbol{\alpha}$ 
  are as follows, 
  \begin{itemize}
    \item $[000]\longmapsto[1100]\longmapsto(\pm a_2,\pm a_2,\pm a_s,\pm a_s)$
    \item $[010]\longmapsto[1000]\longmapsto(\pm a_2,\pm a_1,\pm a_s,\pm a_s)$
    \item $[100]\longmapsto[0100]\longmapsto(\pm a_1,\pm a_2,\pm a_s,\pm a_s)$
    \item $[111]\longmapsto[0000]\longmapsto(\pm a_1,\pm a_1,\pm a_s,\pm a_s)$
    \item $[110]\longmapsto[0000]\longmapsto(\pm a_s,\pm a_s,\pm a_1,\pm a_1)$
    \item $[001]\longmapsto[0011]\longmapsto(\pm a_s,\pm a_s,\pm a_2,\pm a_2)$
    \item $[011]\longmapsto[0010]\longmapsto(\pm a_s,\pm a_s,\pm a_2,\pm a_1)$
    \item $[101]\longmapsto[0001]\longmapsto(\pm a_s,\pm a_s,\pm a_1,\pm a_2)$.
  \end{itemize}
The sign ``$\pm $" of these four output sequences are decided by 
$[c_1c_3c_5c_7]\longmapsto\bs=((-1)^{c_1},(-1)^{c_3},(-1)^{c_5},(-1)^{c_7})\in\{-1,+1\}^4$.

The two 2D projection ouputs of the 7~bit/4D-sym 4D-AC   are also shown in Fig. \ref{fig:mapper_7bit_v2} (right).
As shown in Fig. \ref{fig:mapper_7bit_v2}, 
when X-pol transmits a symbol of $(a_1, a_1)$, a scaled symbol $(a_s, a_s)$ will be transmitted in Y-pol, and vice versa.
The amplitude switch between $a_1$ and $a_s$ is implemented by amplitude scaling operation as red blocks in Fig. \ref{fig:mapper_7bit_v2}, and can be formulated as follows, 
\begin{itemize}
\item If $b_7\oplus {(b_5*b_6)}=0$: $a_1\longmapsto \text{X-pol}$, $ a_s\longmapsto \text{Y-pol}$,
\item If $b_7\oplus {(b_5*b_6)}=1$: $a_s\longmapsto \text{X-pol}$, $a_1\longmapsto \text{Y-pol}$.
\end{itemize}
Note that the white points in Fig. \ref{fig:mapper_7bit_v2} (right) can be considered as the symbols with scaled amplitude of $Ka_2$, which are not selected by amplitude coding.

\section{Numerical Results}

In this section, the performance of the proposed 4D-AC architecture (6b4D-AC and 7b4D-AC) is compared to the original 4D-64PRS and 4D-OS128 formats through numerical simulations. PM-8QAM and 128SP-16QAM are also shown as baselines with the same corresponding spectral efficiencies. The same FEC code rate  $R_c=0.8$ is considered. 4D-64PRS and 4D-OS128 are optimized for the target generalized mutual information (GMI) rate of $4.8$~bit/4D-sym and $5.6$~bit/4D-sym, respectively.

\begin{figure}[!tb]
 \centering
{\includegraphics[]{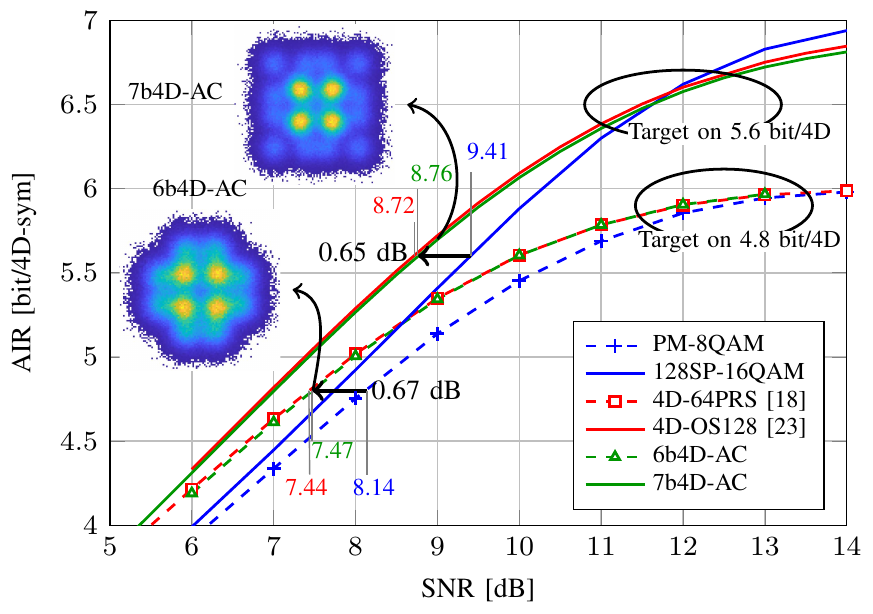}}
\vspace{-2em}  
 \caption{AIR as a function of SNR for various modulation formats. Inset: 2D projection of received constellations.}
\label{fig:GMIvsSNR_AWGN}
\vspace{-1em}
\end{figure}

The results are shown in terms of GMI as an AIR metric for BICM system   
in Fig. \ref{fig:GMIvsSNR_AWGN}. 
The required SNRs for code rate of 0.8 are highlighted by gray lines with the corresponding colored values for different modulation formats in Fig. \ref{fig:GMIvsSNR_AWGN}.
We can observe that two 4D-AC schemes outperform conventional uniform modulation formats by  0.67~dB and 0.65~dB at the rate of 4.8~bit/4D-sym and 5.6~bit/4D-sym, respectively. 
Note that the fully optimized geometrically-shaped 4D-64PRS and 4D-OS128 have 6 and 8 amplitude levels in each real dimension. respectively. By contrast, the 6b4D-AC and 7b4D-AC with 16QAM modulator  only requires 4 and 6 amplitude levels, which can reduce the complexity and resolution requirements of the mapper and demapper. Meanwhile,   the reduced number of amplitude levels may  potentially lead to a small loss of the shaping gain.
The results in Fig. \ref{fig:GMIvsSNR_AWGN}  indicates that 4D-AC based modulation only induce negligible loss ($<0.05$~dB) with respect to  4D-64PRS and 4D-OS128.
Therefore, the performance gains of 4D modulation over PM-8QAM and 128SP-16QAM are preserved.

The 4D symbols are transmitted with the same probability, however, 
due to the imposed  constraints on the MD symbols, the proposed scheme induce a nonuniform probability distribution in each one-dimensional space (see the example  in the right side of  Fig. \ref{fig:structure}) to realize shaping gains. Thus, the received symbols in 2D projection with a different probability are observed in Fig. \ref{fig:GMIvsSNR_AWGN} as insets.

\begin{figure}[!tb]
\centering
 {\includegraphics[]{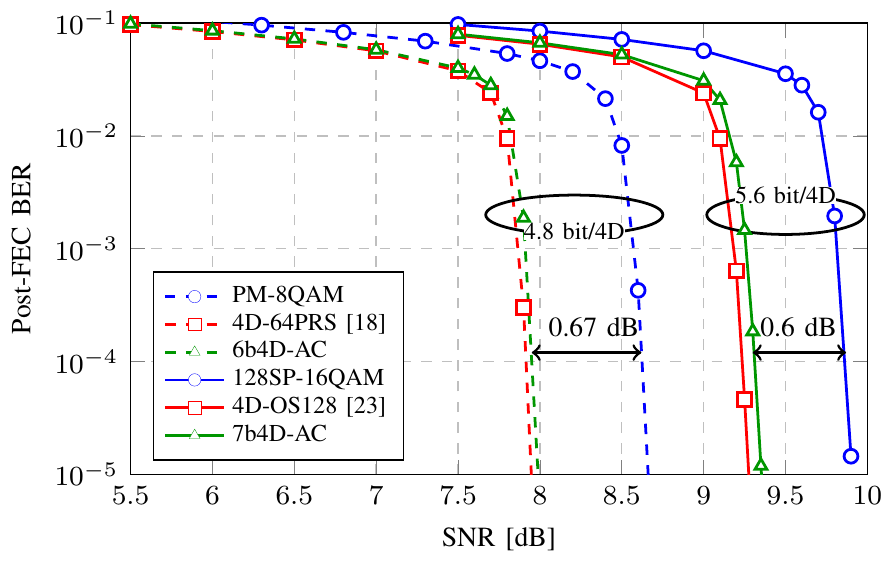}}
 \vspace{-2em}  
 \caption{BER vs. SNR for different modulation formats at transmissions rate of $4.8$~bit/4D-sym and $5.6$~bit/4D-sym, respectively. Rate-$R_c=0.8$ DVB-S2 LDPC codes of length $n_c = 64800$ bits are used. 
 }
\label{fig:postLDPCBER}
\vspace{-1.2em}
\end{figure}

Fig.~\ref{fig:postLDPCBER} shows the decoding performance with the LDPC codes from DVB-S2 standard for 4D-AC modulation formats  
 and uniform signaling (PM-8QAM and 128SP-16QAM) at a transmission rate of 4.8~bits and 5.6~bits per two polarization (bit/4D-sym).
At a BER of $10^{-4}$, the proposed 4D-AC shaping scheme outperforms conventional modulation formats PM-8QAM and 128SP-QAM by approximately up to 0.67~dB.

\section{Conclusions}
We have proposed to use multidimensional amplitude coding (MD-AC) to realize geometrical shaping. The proposed architecture only requires simple logic circuit operations. Implementations for formats with 6~bit/4D-sym and 7~bit/4D-sym were presented. The proposed solution is expected to reduce the mapper complexity with negligible performance loss with respect to fully optimized geometrically-shaped 4D formats {and can be extended to higher dimensional formats with higher spectral efficiencies.}

While probabilistic shaping offers in principle larger gains than geometrical shaping (and rate adaptivity), its theoretically superior performance in terms of rates can vanish due to rate loss in the finite blocklength regime and nonlinearity penalties \cite[Fig. 10]{ChenJLT2021}. The proposed geometric shaping method can be considered as a competitive shaping approach with low complexity to increase the performance of optical communication systems.

\bibliographystyle{IEEEtran}
\bibliography{references_4D64PRS,references,ref}

\end{document}